\documentclass[a4paper]{jpconf}
\pdfoutput=1
\usepackage{graphicx}

\begin{document}

\title{Dimensional crossover in spin-1 Heisenberg antiferromagnets: a quantum Monte Carlo study}
\author{Keola Wierschem and Pinaki Sengupta}
\address{Division of Physics and Applied Physics, School of Physical and Mathematical Sciences, Nanyang Technological University, 21 Nanyang Link, Singapore 637371}
\ead{keola@ntu.edu.sg, psengupta@ntu.edu.sg}

\begin{abstract}
We present results of large scale simulations of the spin-1
Heisenberg antiferromagnet on a tetragonal lattice.
The stochastic series expansion quantum Monte Carlo method is used
to calculate equilibrium thermodynamic variables
in the presence of an external magnetic field.
In particular, the low temperature magnetization curve
is investigated in the quasi-one-dimensional (Q1D),
quasi-two-dimensional (Q2D), and three-dimensional (3D) limits.
Starting from the 3D limit,
the Q1D (Q2D) limit is achieved by reducing the in-plane (out-of-plane)
spin coupling strength towards zero.
In the Q1D limit, a Haldane gap appears in the magnetization curve
at low magnetic field.
Additionally, near the saturation field the slope of the magnetization
curve increases substantially, approaching the infinite-slope behavior
of a one-dimensional spin-1 chain. A similar (though less pronounced)
effect is seen in the Q2D limit.
We also study the effect of uniaxial single-ion anisotropy on the
magnetization curve for Q1D and Q2D systems.
Our results will be important in understanding the field-induced
behavior of a class of low-dimensional Ni-based quantum magnets.
\end{abstract}

\section{Introduction}

The nature of the ground state phases in interacting quantum many body systems is often strongly influenced by the spatial dimensionality. This is a direct consequence of enhanced quantum fluctuations that result in novel quantum phases  in systems with reduced dimensionality that are not present in the isotropic limit. A prominent example of this phenomenon is the Haldane gap phase in spin-1 Heisenberg antiferromagnetic (HAFM) chains. Haldane proved that the ground state of the one-dimensional spin-1 HAFM is gapped~\cite{haldane1983} whereas there is strong numerical evidence that the ground state in higher dimensions is the gapless Ne{\' e}l state~\cite{lin1989}.

In this paper, we study the emergence of the Haldane phase as the quasi-one-dimensional (Q1D) limit is approached, starting from the isotropic spin-1 HAFM and varying the interchain coupling continuously. The quasi-two-dimensional (Q2D) limit is also explored by varying the inter-layer coupling in a continuous manner. The zero-field 1D phase diagram has been well mapped~\cite{chen2003}, and the 1D field-dependence has also been reported~\cite{sengupta2007}. The 2D phase diagram is also well known~\cite{hamer2010}. Here, we attempt to extend this knowledge to the Q1D and Q2D limits, and to make a connection to the isotropic 3D limit. In addition to elucidating the nature of the emergence of the effects of reduced dimensionality, our results have direct relevance to recent experiments on Q1D and Q2D Ni-based spin-1 Heisenberg quantum antiferromagnets.

\section{Theory}

The Hamiltonian for a spin-1 Heisenberg antiferromagnet on an anisotropic cubic lattice is
\begin{equation}
H=J_{\perp}\sum_{\left<ij\right>_{\perp}}\vec{S_{i}}\cdot\vec{S_{j}}
+J_{\parallel}\sum_{\left<ij\right>_{\parallel}}\vec{S_{i}}\cdot\vec{S_{j}}
-h\sum_{i}S_{i}^{z}
+D\sum_{i}\left(S_{i}^{z}\right)^{2},
\end{equation}
where $h$ is the applied magnetic field, $D$ is a single-ion anisotropy parameter, and $\left<ij\right>_{\perp}$ ($\left<ij\right>_{\parallel}$) represents nearest neighbor spin pairs with in-plane (out-of-plane) bonds. The saturation field can be determined by a zero energy cost to flip a single spin in the fully polarized ground state, and is given by $h_{s}=D+4(2J_{\perp}+J_{\parallel})$. We define a dimensional anisotropy parameter $r$ as the ratio $J_{\perp}/J_{\parallel}$ ($J_{\parallel}/J_{\perp}$) for Q1D (Q2D) systems. Thus, in either the Q1D or Q2D case, the low-dimensional limit is achieved as $r$ approaches zero.

We have performed large-scale numerical calculations of the above Hamiltonian using the stochastic series expansion quantum Monte Carlo method with operator-loop updates~\cite{sandvik1999}. In all our simulations, either $J_{\parallel}$ or $J_{\perp}$ is always set to 1, so that the inverse temperature, $\beta$, is in units relative to the isotropic interaction strength ($J_{\parallel}$ in the Q1D regime and $J_{\perp}$ in the Q2D regime).

\section{Results}

\begin{figure}
\includegraphics[clip,trim=0 3in 0 0.25in,width=0.5\linewidth]{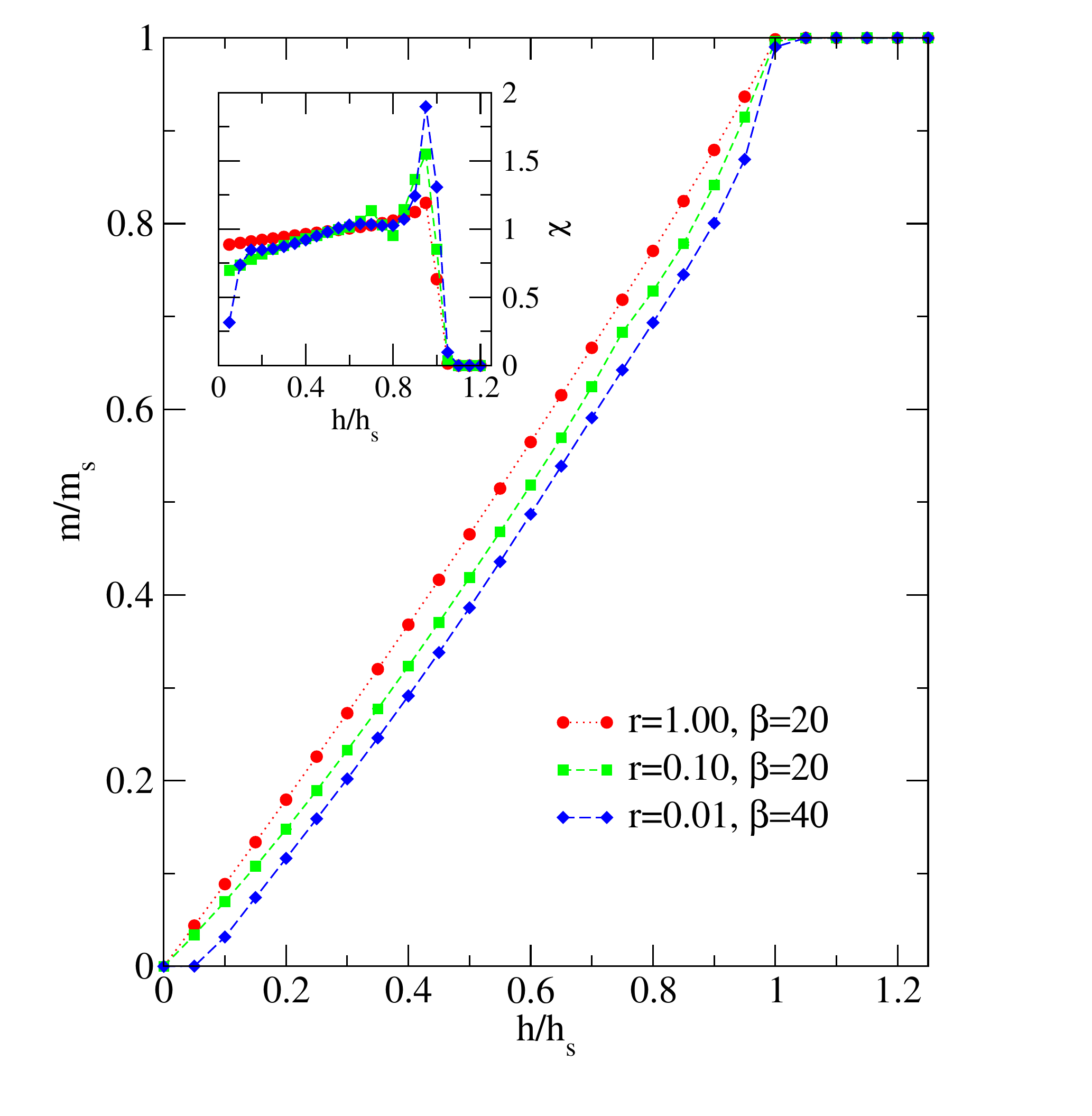}
\includegraphics[clip,trim=0 3in 0 0.25in,width=0.5\linewidth]{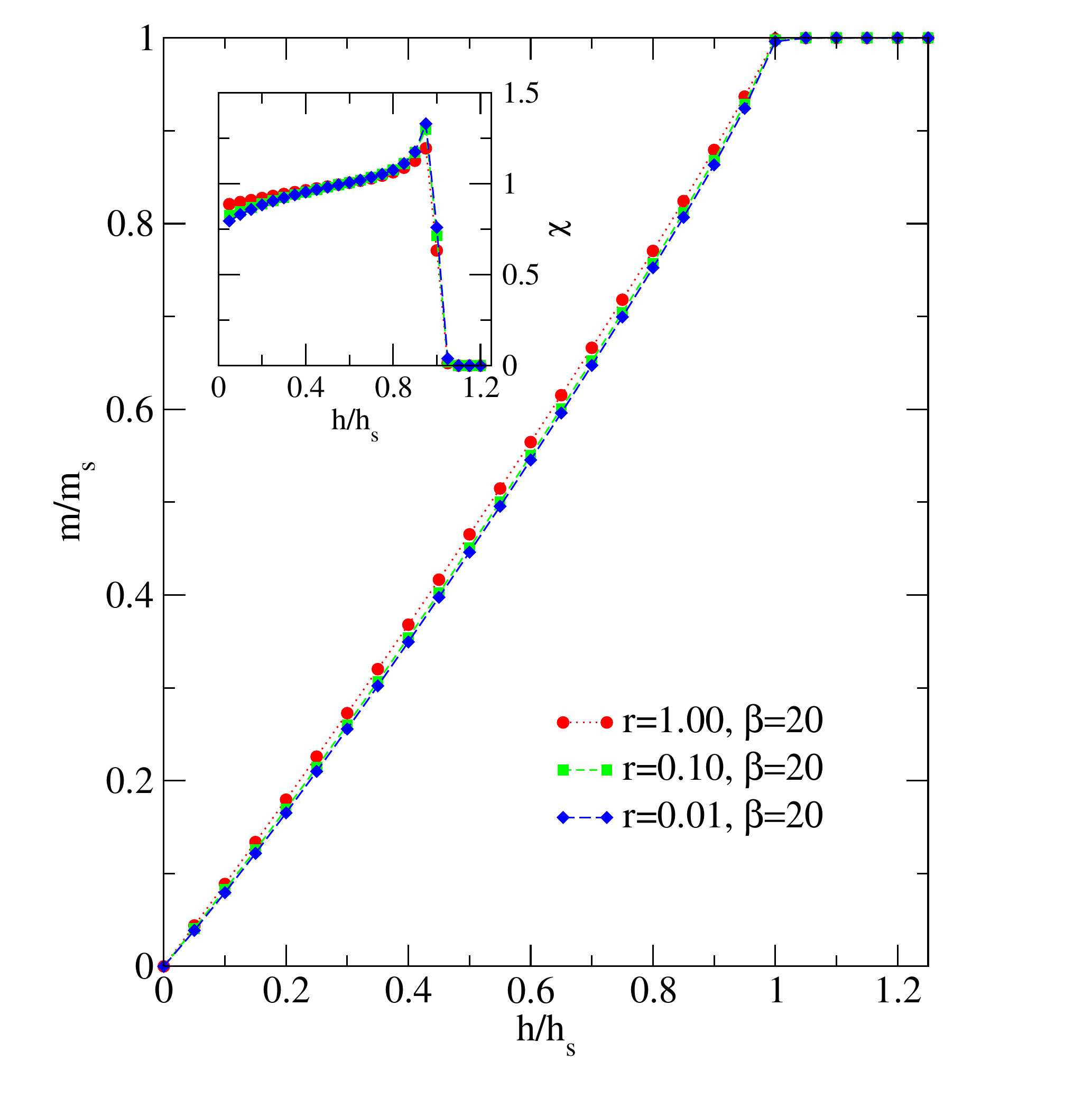}
\caption{\label{1d2d}Magnetization and susceptibility approaching the Q1D (left) and Q2D (right) limits. Results are given for tetragonal lattices of $8\times8\times32$ (Q1D) and $16\times16\times8$ (Q2D).}
\end{figure}

In Figure~\ref{1d2d} we show the magnetization curves as we approach the Q1D (left panel) and Q2D (right panel) limits. Shown in the insets is the uniform static susceptibility, $\chi$, which displays a peak just before the saturation field. The height of this peak increases as $r$ is diminished, the effect being most pronounced for the Q1D system. Additionally, for the Q1D system with $r=0.01$, a Haldane gap appears in the magnetic response at low fields.

\begin{figure}
\begin{minipage}{0.47\linewidth}
\includegraphics[clip,trim=0 3in 0 0.25in,width=\linewidth]{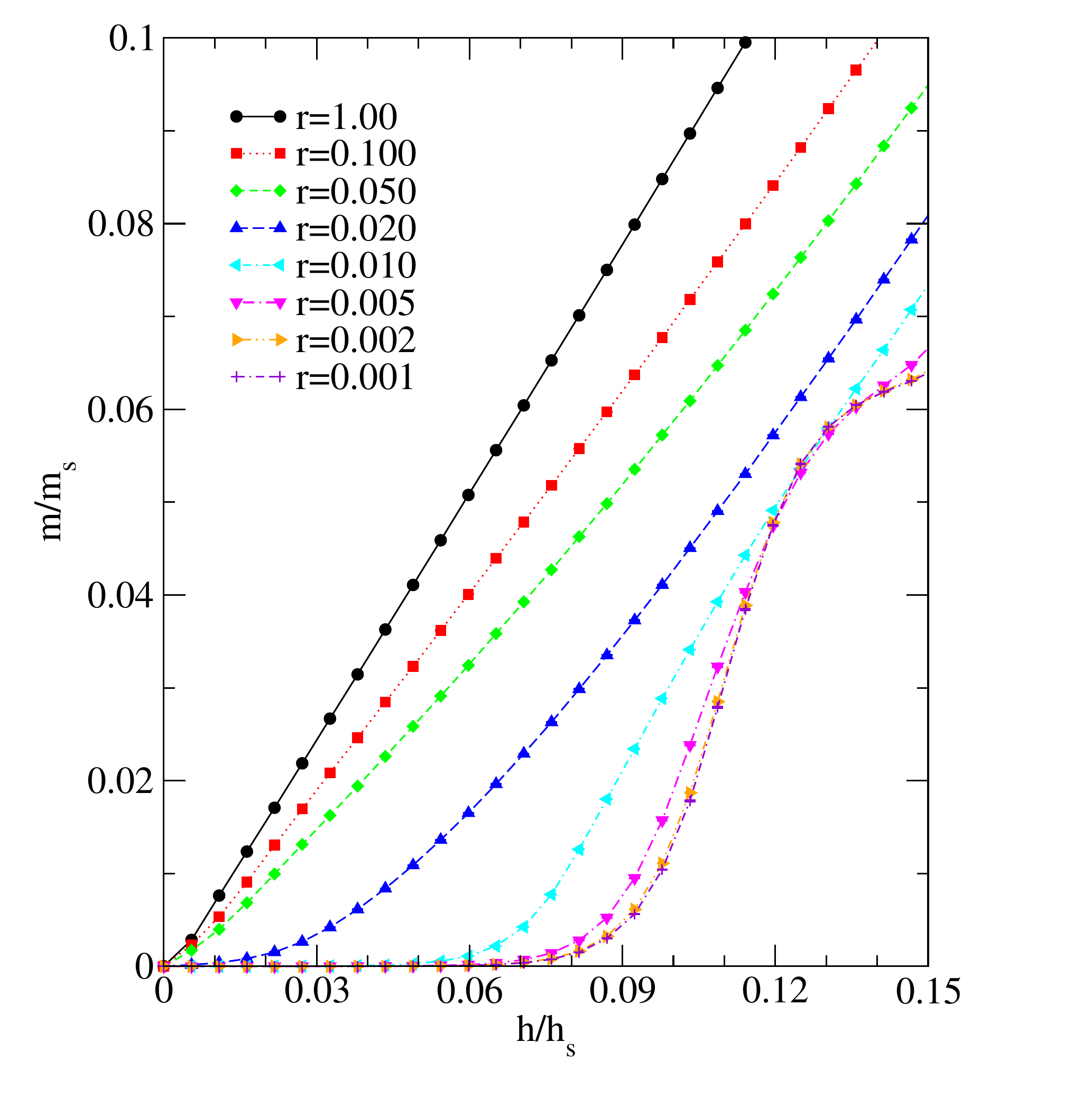}
\caption{\label{crossover}Magnetization at low field for the Q1D system at $\beta=32$. A Haldane spin-gapped phase emerges for $r\approx0.02$ and below. The plateau at $m/m_{s}=1/16$ is due to the finite size of our system ($4\times4\times16$).}
\end{minipage}
\hspace{0.05\linewidth}
\begin{minipage}{0.47\linewidth}
\includegraphics[clip,trim=0 3in 0 0.25in,width=\linewidth]{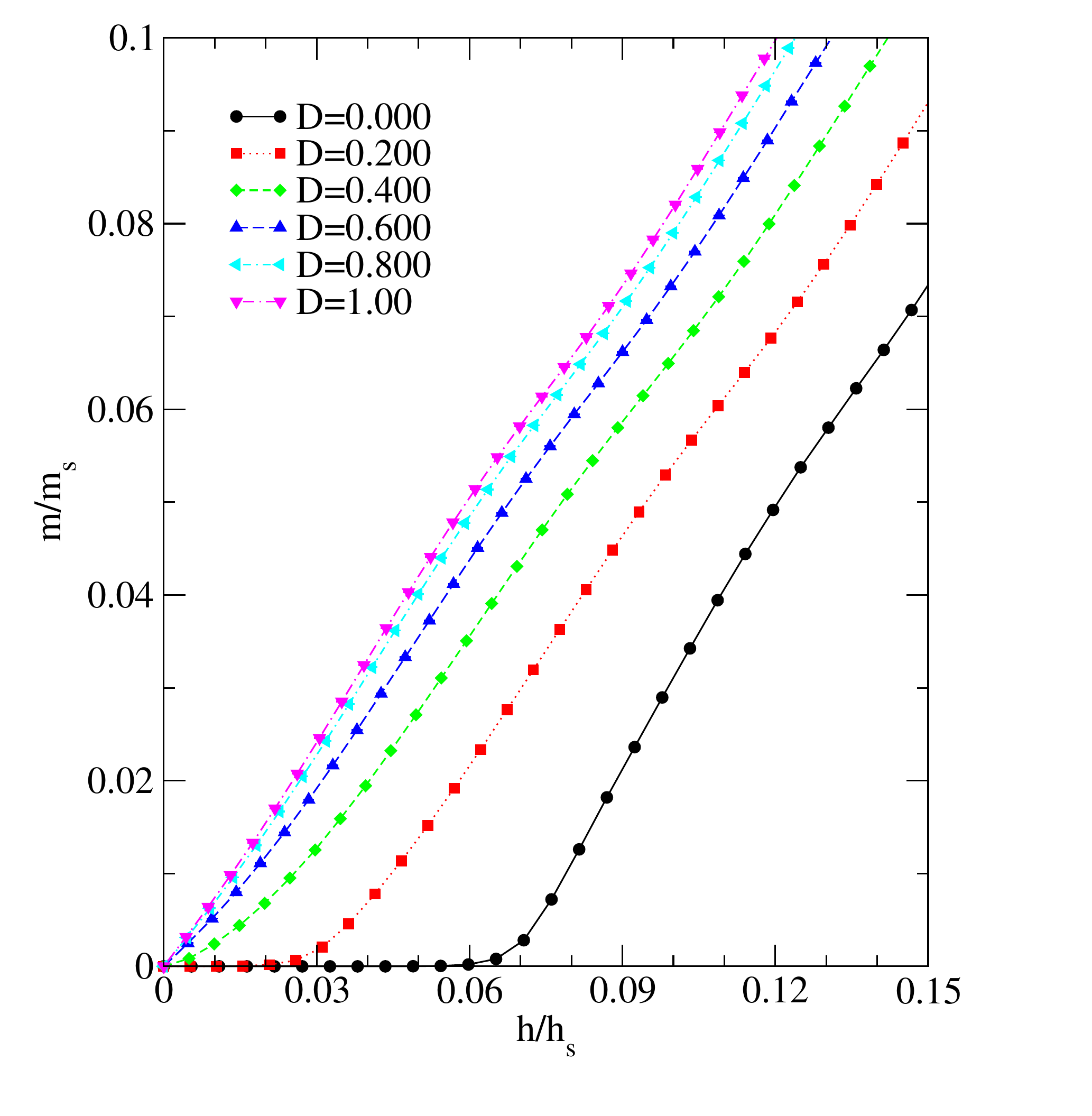}
\caption{\label{dd}Magnetization at low field for the Q1D system with $r=0.01$. As the easy-plane anisotropy $D$ is increased to $0.40$ and above, the Haldane gap closes. Results are obtained on a $4\times4\times16$ lattice at $\beta=64$.}
\end{minipage}
\end{figure}

Next, we turn our attention to the low-field behavior of the Q1D systems, where a spin-gapped Haldane phase is expected for 1D systems~\cite{haldane1983}. In Figure~\ref{crossover} we plot the uniform magnetization at low field for systems ranging from the isotropic 3D limit to the Q1D limit. In the extreme Q1D limit ($r<0.01$), a spin gap is clearly seen in the magnetic response to applied field. In the other limit for $r$ near the isotropic 3D value, no gap is present. Thus, we see a crossover from 3D to 1D behavior at a critical value of the anisotropy, $r_{c}$, that we estimate to be $\approx0.02$, as this value separates the two observed limits. Whether or not there is a Haldane gap precisely at $r=0.02$ is difficult to determine due to possible finite-size and finite-temperature effects.

Now we determine the effect of adding an easy-plane anisotropy to the zero-field Q1D system in the effective 1D limit ($r=0.01$). The Haldane gap phase is destroyed as the value of $D$ is increased, as shown in Figure~\ref{dd}. Here we note that this occurs rather rapidly, as no gap is seen for $D>0.2$, when it is known that in 1D the gap persists up to $D\approx1$~\cite{sakai1990}.

\begin{figure}
\includegraphics[clip,trim=0 3in 0 0.25in,width=0.5\linewidth]{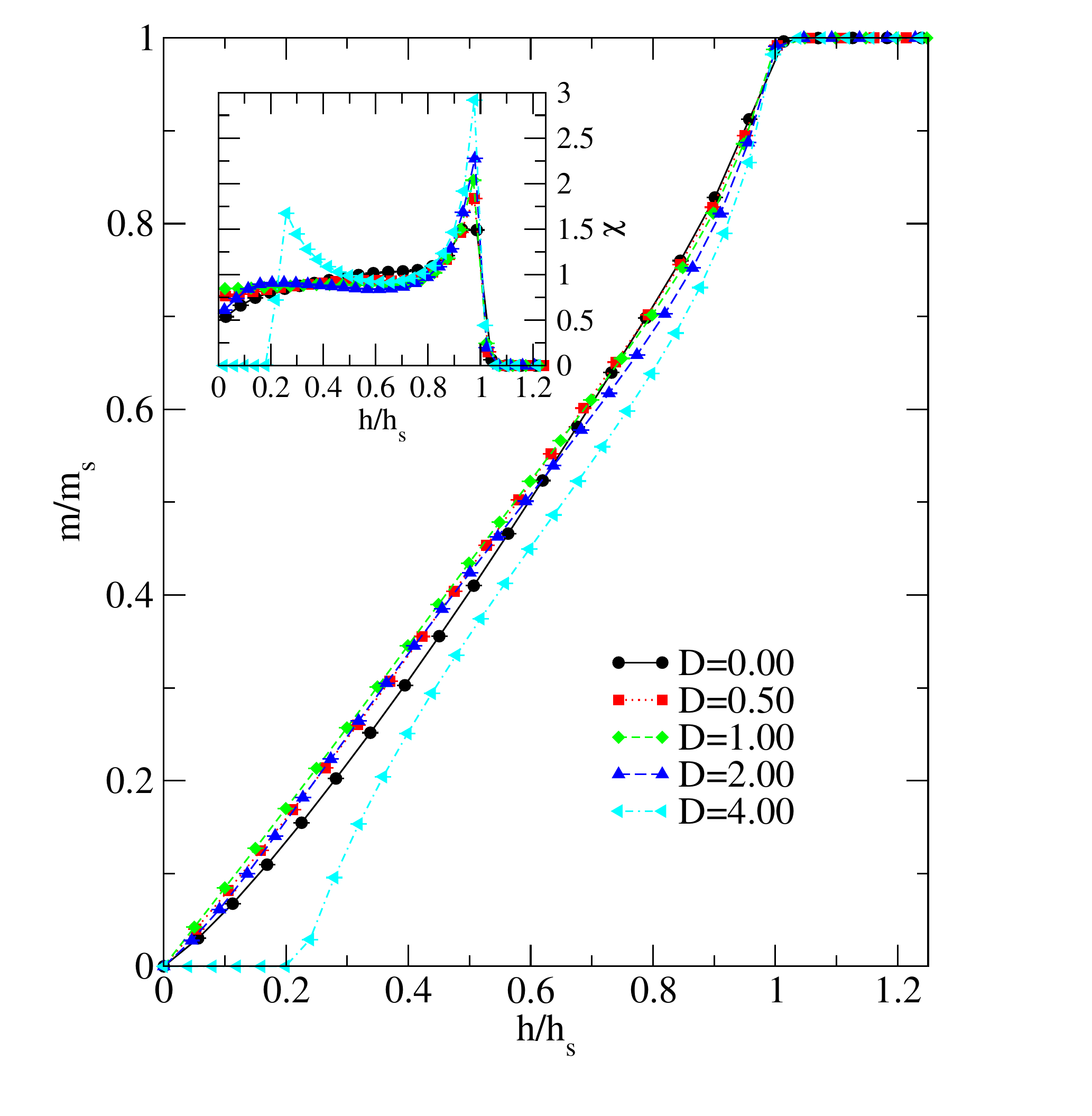}
\includegraphics[clip,trim=0 3in 0 0.25in,width=0.5\linewidth]{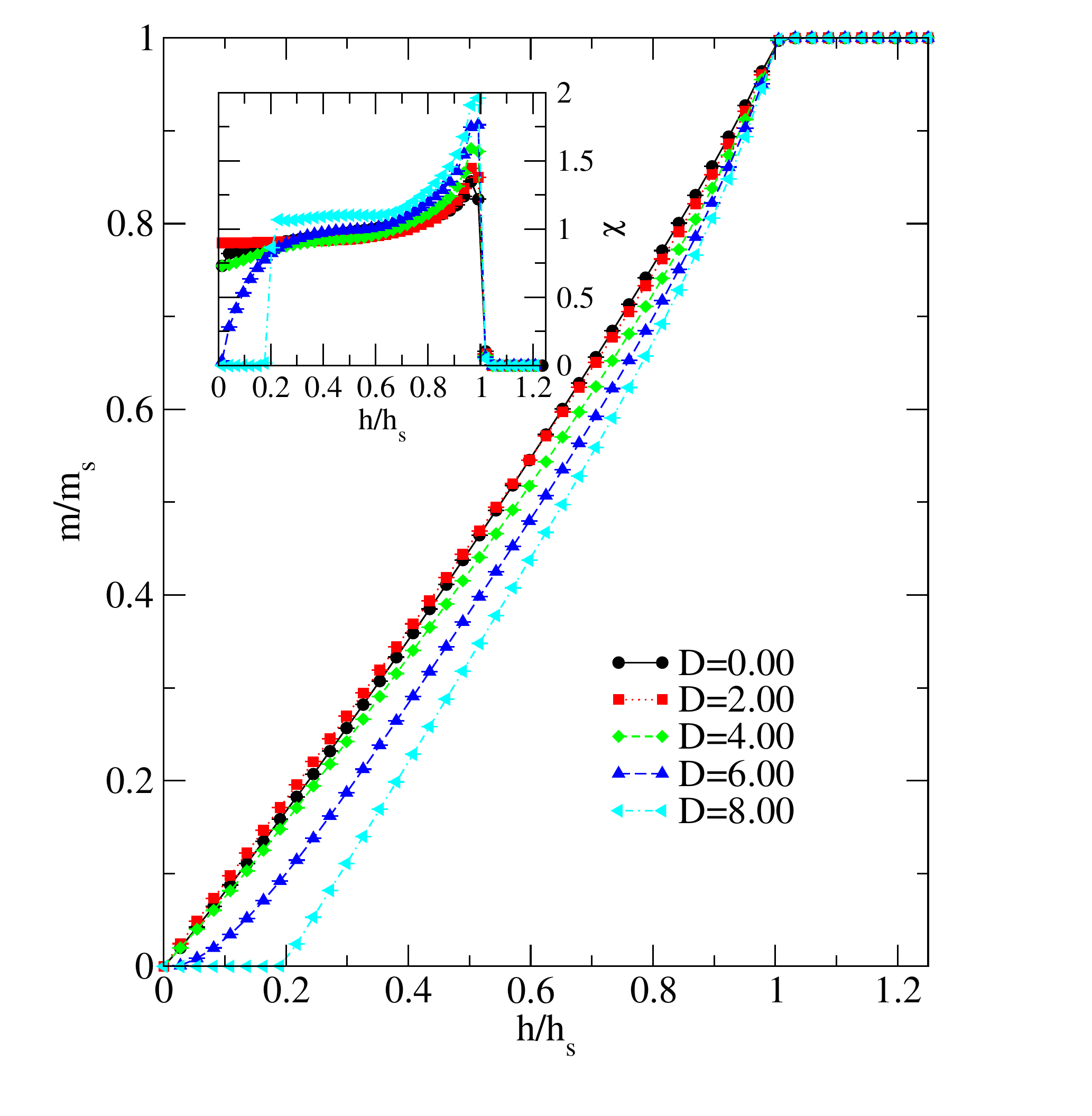}
\caption{\label{1dd2dd}Magnetization and susceptibility for $D\ge0$ in the Q1D (left) and Q2D (right) limits. Results are obtained on $4\times4\times16$ (Q1D) and $8\times8\times4$ (Q2D) lattices at $\beta=16$ and $r=0.05$.}
\end{figure}

Finally, we show the full field behavior for $D\ge0$ in the Q1D and Q2D limits in Figure~\ref{1dd2dd}. We find that the peak in $\chi$ near the saturation field increases with increasing $D$ for both Q1D and Q2D systems. A broad, shallow peak also emerges in the Q1D system for $D=2.00$. This may be associated with the approach of a quantum paramagnetic phase at large $D$. This spin-gapped phase is seen in the Q1D system at $D=4.00$ and in the Q2D system for $D=8.00$.

\section{Discussion}

In the Haldane phase at zero field, there is a spin gap $\Delta$ to magnetic excitations. This gap closes under an applied field $h_{c}=\Delta$~\cite{maeda2007}. Below this critical value of the magnetic field the ground state magnetization is zero. In 1D the Haldane gap is well known to be $\Delta=0.41050(2)$, in units of $J_{\parallel}$~\cite{white1993}. Thus $h_{c}\approx0.4J_{\parallel}$ or, utilizing $h_{s}=4J_{\parallel}$ in 1D, $h_{c}\approx0.1h_{s}$. We can estimate $h_{c}$ for our Q1D calculations from Figure~\ref{crossover}. Taking into account thermal effects, we find $h_{c}\approx0.09h_{s}$ in the effective 1D limit. This is slightly below, but in general agreement with, the pure 1D result.

We can also compare our estimate of $r_{c}$ ($\approx0.02$, see Figure~\ref{crossover}) with a mean field theory (MFT) result of $\approx0.013 $~\cite{sakai1990}. The difference between our value and the MFT calculation can be attributed to two likely causes. One is that we are using relatively small system sizes, which will tend to show finite size gaps. The other is a tendency for the MFT to overestimate order, implying that $r_{c}$ from MFT may be a lower bound~\cite{sakai1990}.


\section{Conclusions}

We have shown a crossover from 3D to Q1D physics as the in-plane spin couplings are reduced towards zero (relative to the easy-axis spin couplings). The result is the emergence of the Haldane phase for $r<r_{c}\approx0.02$. The behavior of the magnetization curve in the Q2D limit was also calculated. The susceptibility peak near the saturation field was found to increase with decreasing $r$. This same effect appears in the Q1D limit, though much more pronounced. We also showed the effect of an easy-plane single-ion anisotropy. In addition to sharpening the susceptibility peak, a gapped phase appears for large values of $D$.

These results are relevant to understanding the magnetization processes in real materials. In particular, the gapless states of the Q1D and Q2D systems are expected to help characterize the magnetic behavior of a class of low-dimensional Ni compounds. We have already seen promising qualitative agreement with our Q2D simulations.

\section*{Acknowledgments}

We would like to thank Paul Goddard for stimulating our interest in this work as well as for discussions regarding the  experimental relevance of this model to Ni-based quantum magnets.

\end{document}